\documentclass[aps,amsmath,amssymb,twocolumn,prl,superscriptaddress]{revtex4-2}

\usepackage{graphicx}
\usepackage{dcolumn}
\usepackage{bm}
\usepackage{color}
\usepackage[normalem]{ulem}
\usepackage{amsmath}
\usepackage{hyperref}

\usepackage{xcolor}
\usepackage{cancel}
\usepackage{braket}
\usepackage{calc}
\usepackage{bbold}
\usepackage{bm}
\usepackage{ulem}

\begin{document}

\title{Thermodynamic uncertainty relations in superconducting junctions}

\author{David Christian Ohnmacht} 
\affiliation{Fachbereich Physik, Universität Konstanz, D-78457 Konstanz, Germany}
\author{Juan Carlos Cuevas}
\affiliation{Departamento de F\'{\i}sica Te\'{o}rica de la Materia Condensada,
Universidad Aut\'{o}noma de Madrid, E-28049 Madrid, Spain}
\affiliation{Condensed Matter Physics Center (IFIMAC), Universidad Aut\'{o}noma de Madrid, E-28049 Madrid, Spain}
\author{Wolfgang Belzig} 
\affiliation{Fachbereich Physik, Universität Konstanz, D-78457 Konstanz, Germany}
\author{Rosa L\'{o}pez}
\affiliation{Institute for Cross-Disciplinary Physics and Complex Systems IFISC (UIB-CSIC), E-07122 Palma de Mallorca, Spain}
\author{Jong Soo Lim}
\affiliation{Department of molecular design, Arontier Co., 15F Sewon Bldg., Gangnam-daero 241, Seocho-gu, Seoul, Republic of Korea}
\author{Kun Woo Kim}
\affiliation{Department of Physics, Chung-Ang University, 06974 Seoul, Republic of Korea}

\date{\today}

\begin{abstract}
It has been recently predicted that the thermodynamic uncertainty relation (TUR),
which expresses a trade-off between the amount of dissipation and the absence of fluctuations, can be 
violated in the context of charge transport in normal quantum conductors. However, the required conditions
are difficult to meet and no experimental violation has been yet reported in mesoscopic quantum transport 
experiments. Here, we predict that TUR can be largely violated in superconducting contacts without 
introducing an energy dependency in the normal transmission coefficient. This violation originates from 
the coexistence of tunneling processes involving different numbers of transmitted charges, namely 
quasiparticle tunneling and Andreev reflections. The superconducting junctions invoked here have been 
fabricated by many groups and constitute an ideal platform to demonstrate TUR's breaking in charge transport 
experiments.
\end{abstract}

\maketitle

\emph{Introduction}.---
The implementation of superconducting materials has become a key ingredient to design better and more functional 
quantum thermal machines. They have been proposed to be efficient for cooling performances 
\cite{nahum1994electronic,leivo1996efficient,PhysRevB.98.241414,PhysRevB.107.245412,PhysRevResearch.5.043041}, 
as well as for pumping or generating power \cite{PhysRevB.106.115419,PhysRevB.103.235434}. Besides, superconductors 
can exhibit remarkable thermoelectric properties in the nonlinear transport regime when electron-hole symmetry 
is broken \cite{PhysRevLett.124.106801,germanese2022bipolar,PhysRevResearch.6.L012022,PhysRevB.106.245413}.

Another advantageous aspect when dealing with superconducting elements is that they overcome the fundamental 
bound imposed by the thermodynamic uncertainty relation (TUR) 
\cite{Barato2015,Todd2016,Horowitz2020,Pietz2016,Pietzonka16b,Pietz2018,Timp2019,PhysRevB.98.155438,PhysRevResearch.5.043041}. 
The TUR establishes a trade-off relation between uncertainty of a current traversing the system and 
the associated entropy production or dissipation. The TUR is a nonequilibrium relation derived from the Markovian 
evolution of a system and sets a more restrictive bound for the entropy production in terms of the noise to current 
ratio. In the case of charge transport, TUR dictates $S/ I^2\geq 2k_{\rm B}/\dot\Sigma$ with $k_{\rm B}$ the Boltzmann 
constant, $\dot\Sigma$ the entropy production, $I$ the average charge current and $S$ the corresponding current fluctuations.

TUR violations have been predicted in quantum systems attributed to either quantum coherence or 
the breakdown of the local detailed balance \cite{Andrieux_2007,PhysRevLett.120.090601,PhysRevX.11.021013,
PhysRevResearch.5.043041,PhysRevResearch.5.013038,PhysRevResearch.5.023155,PhysRevB.28.1655}. Large deviations from 
the TUR are desirable for the quantum-enhanced performance of the engine operation, i.e., less dissipative and more 
stable machines \cite{PhysRevB.98.085425,PhysRevB.98.155438,PhysRevLett.120.090601,PhysRevE.99.062141}. Besides, quite 
recently there has been a notable effort towards finding quantum systems exhibiting a prominent breakdown of the TUR
\cite{PhysRevResearch.5.043041,PhysRevResearch.5.023155,PhysRevB.104.045424,PhysRevResearch.5.013038,PhysRevResearch.5.023155,PhysRevB.108.115422}. 
Thus, for instance, a chain of quantum dots with a rectangular transmission was proposed to violate the standard TURs 
\cite{PhysRevB.104.045424}. Additionally, moderate deviations in the TUR relation have been predicted 
in hybrid normal-superconductor quantum dots in the weak tunnel coupling regime \cite{PhysRevResearch.5.043041}.
However, all these proposals require stringent conditions and all attempts to 
observe a violation of TUR in quantum conductors have failed thus far, see e.g.\ Ref.~\cite{PhysRevB.101.195423}.

Here, we predict that TUR can be largely violated in few-channel quantum point contacts involving
superconducting leads, which have already been realized e.g.\ with the help of superconducting atomic-size contacts. 
We show that the strong violations of TUR in these junctions are due to the coexistence of different transport processes,
namely single-quasiparticle tunneling (QP) and (multiple) Andreev reflections (AR). We find that the largest TUR
violations occur for highly transmissive contacts and no energy dependence of the normal state transmission is required, 
contrary to what happens in normal quantum point contacts \cite{PhysRevB.98.085425,PhysRevB.104.045424}.

\emph{Quantifying TUR deviations.---} In what follows we present a general discussion to explain
what makes superconductors special in the context of TUR. For this purpose, we consider a generic isothermal two-terminal 
quantum device that may contain normal metals (N) and/or superconductors (S). If we apply a bias voltage $V$, entropy 
production due to the Joule heating is given by $\dot \Sigma=I V/ T$, where $I$ is the charge current and $T$ the temperature
of the electrodes. Deviations for the TUR, i.e., $\frac{S}{I^2}\dot\Sigma\geq 2k_B$ with $S$ being the current fluctuations, 
can be quantified by the TUR-breaking coefficient
\begin{eqnarray}\label{TURcoeff}
\mathcal{F}\equiv F^*-\frac{2k_{\rm B}T}{V},
\end{eqnarray}
where $F^* = S/I$ is the Fano factor \textcolor{black}{and we have assumed that the electron charge is unity.} 
The coefficient $\mathcal{F}$ is negative for broken TUR and positive (or zero) for unbroken TUR, where 
$\mathcal{F}= 0$ corresponds to the classical limit. \textcolor{black}{Below, we shall present a quantitative 
analysis of $\mathcal{F}$ valid for arbitrary range of parameters, but to get an insight into the conditions 
required to break TUR, we now follow Ref.~\cite{PhysRevB.98.085425} and focus on the low-bias regime in which
it is possible to make analytical progress. With this idea in mind, we make a Taylor expansion in the bias 
of the current and noise}
\begin{eqnarray}
I & = & G_1 V + \frac{1}{2!} G_2V^2 + \frac{1}{3!} G_3V^3 + \mathcal{O}(V^4), \\
S & = & S_0 + S_1 V + \frac{1}{2!} S_2 V^2 + \mathcal{O}(V^3) . \quad\quad
\end{eqnarray}
Here, $G_1, G_2, G_3$ are linear and nonlinear conductances and $S_0, S_1, S_2$ are the equilibrium 
and nonequilibrium noise terms.  Using the nonlinear fluctuation relations \cite{PhysRevLett.101.136805,PhysRevB.72.235328,
Andrieux_2007,PhysRevLett.108.246603} together with the Johnson-Nyquist relation for the equilibrium noise 
(i.e., $S_0 = 2k_{\rm B}T G_1$), we obtain $\mathcal{F} = Vk_{\rm B}T/(G_1)C_{\rm neq} + \mathcal{O}(V^2)$ where the 
coefficient \cite{PhysRevB.98.155438}
\begin{equation}
    C_{\text{neq}} \equiv \frac{3S_2 - 2k_{\rm B} T G_3}{6k_{\rm B}T}
    \label{Cneq}
\end{equation}
quantifies TUR deviations. Namely when $C_{\text{neq}}< 0$ (in the limit of small voltages), TUR is violated.

The stochastic nature of charge transfer in mesoscopic junctions allows us to express the current in an arbitrary 
mesoscopic junction as follows \cite{Levitov1996,Klich2003}
\begin{equation}\label{current}
I = \int \frac{dE}{h} \left(\sum_n n p_n\right),
\end{equation}
where $p_n = p_n(E,V,T)$ are charge-resolved probabilities of a transport process transferring $n$ charges across 
the junction that depends on the energy $E$, voltage $V$ and temperature $T$. The corresponding 
current fluctuations are given by \cite{Levitov1996,Klich2003}
\begin{equation}\label{current_noise}
    S = \int \frac{dE}{h} \left[ \sum_n n^2 p_n -\left(\sum_n np_n\right)^2 \right].
\end{equation}
These formulas for $I$ and $S$ hold for junctions containing normal metals and/or superconductors. 
For normal contacts, the transport is solely due to single-electron tunneling ($|n|=1$), while with S electrodes 
multiple electron ($|n|>1$) tunneling processes can coexist. This is what makes superconductors special, as we are about to show.  

To get further insight, we expand the charge-resolved probabilities in the applied voltage
\begin{equation}\label{eq:prob}
    p_n = p_n^{(0)}+p_n^{(1)}V+p_n^{(2)} V^2/2+p_n^{(3)}V^3/6+ \mathcal{O}(V^4).
\end{equation}
By using the detailed balance condition $  p_{-n} = p_n e^{-n\beta V}$ \cite{PhysRevB.72.235328} where 
$\beta = 1/(k_{\rm B}T)$ and inserting the expanded probabilities [see Eq.~(\ref{eq:prob})] in the current and noise 
formula [see Eqs.~(\ref{current}), and (\ref{current_noise})] we obtain the following result for the coefficient 
$C_{\rm neq}$ in Eq.~(\ref{Cneq})
\begin{align}\label{deviation}
    \frac{6 C_{\rm neq}}{\beta^3} = \int \frac{dE}{h} \Bigg ( & \sum_{n>0} n^4p_n^{(0)}(1-6p_n^{(0)}) \nonumber \\ 
    & -\sum_{m>n>0} 12n^2m^2 p_n^{(0)}p_m^{(0)} \Bigg ),
\end{align}
which notably only depends on the zeroth order approximation of the transmission probabilities $p_n^{(0)}$.

\begin{figure}
    \centering
    \includegraphics[width=0.5\textwidth,clip]{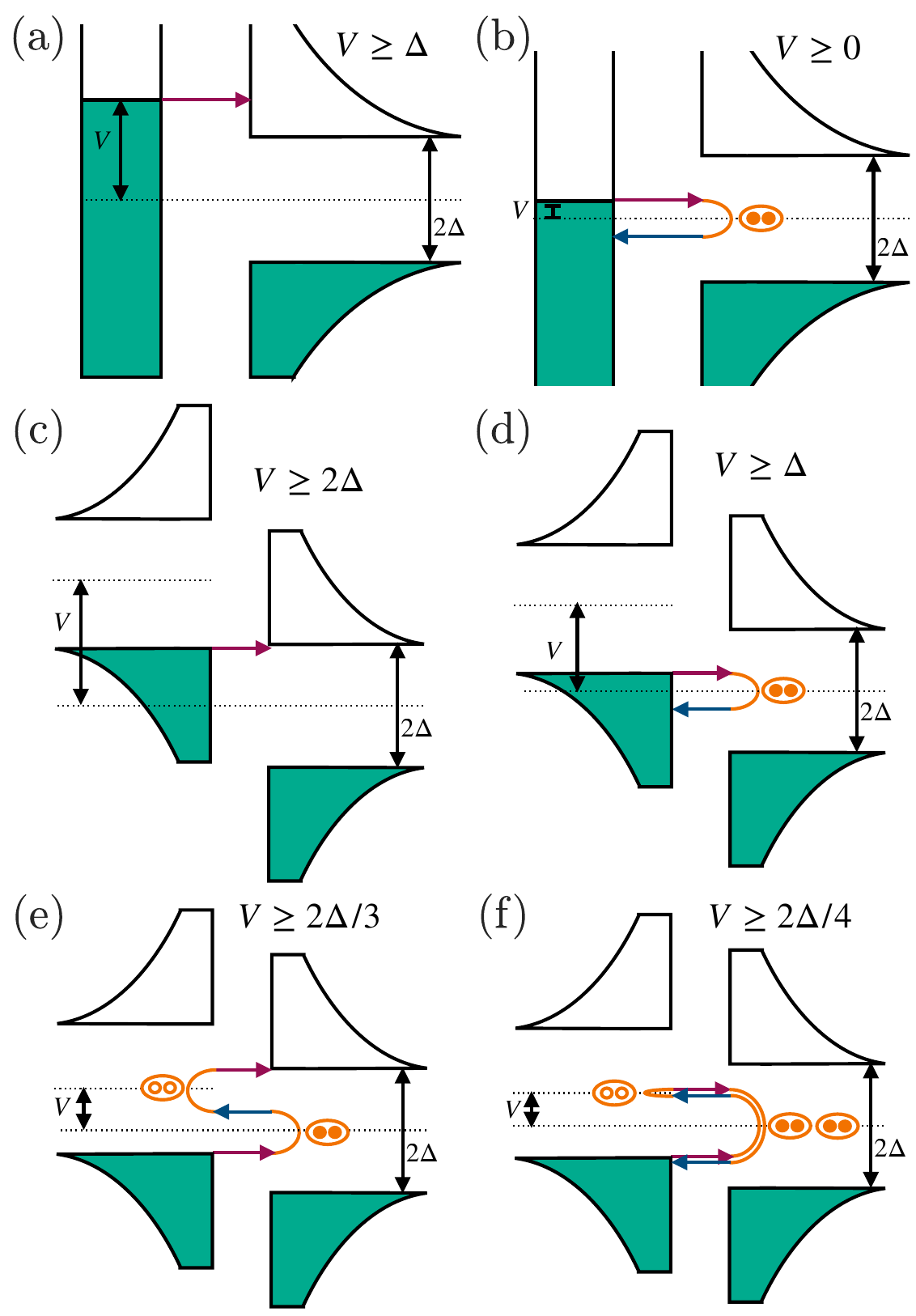}
    \caption{(a) Single-QP tunneling in which a QP tunnels into the empty density of states of the superconductor 
    transferring one electron. The applied voltage has to exceed the SC gap, $V \geq \Delta$. (b) AR in which an electron 
    is reflected as a hole inside the superconducting gap transferring two electrons. The voltage is arbitrary, $V\geq 0$. 
    (c) QP tunneling between two SCs at voltages above $2\Delta$. (d) AR from a voltage biased superconductor to another
     superconductor transferring two electrons. The voltage has to exceed the SC gap, $V \geq \Delta$. (e) First MAR, 
    where an incoming electron is reflected as a hole and that hole is retro-reflected as another electron, transferring 
    a total of three electrons. The voltage has to exceed $V \geq 2\Delta/3$. (f) Higher order MAR involves a process that 
    transfers four electrons with onset voltage $V \geq 2\Delta/4$. The voltage thresholds correspond to the case of zero 
    temperature.}
    \label{fig-1}
\end{figure}

To exemplify Eq.~(\ref{deviation}) in the context of superconducting junctions, we now consider a single-channel NS 
contact. The superconducting lead is assumed to be a conventional BCS superconductor with a temperature-dependent 
gap $\Delta \equiv \Delta(T)$. In this case, there are two transport processes as illustrated in Fig.~\ref{fig-1}(a,b). 
For zero temperature, QPs can only be transferred for energies above the gap ($V\geq \Delta$) and transfer one charge 
which is shown in panel Fig.~\ref{fig-1}(a). In the subgap region ($V \leq \Delta$) transport can take place via Andreev
reflection where two charges are transferred in one instance, see panel Fig.~\ref{fig-1}(b). For non-zero temperatures, 
QP transfer can also occur for in-gap voltages because of the thermal broadening of the corresponding Fermi functions. 
The two processes are associated with different transmission probabilities, namely $p_{\pm1}$ corresponds to QP tunneling 
and $p_{\pm 2}$ to AR and all other probabilities are zero \cite{Belzig2003}. With this information at hand, the coefficient
$C_{\rm neq}$ reads:
\begin{equation}\label{CneqNS}
    \frac{6C_{\rm neq}^{\rm NS}}{\beta^3}{=}\int \frac{dE}{h} 
    \left[p_1^{(0)}(1{-}6p_1^{(0)}){+}2^4p_2^{(0)}(1{-}6p_2^{(0)}){-}48p_1^{(0)}p_2^{(0)}\right].
\end{equation}
This expression reduces to the result in Ref.~\cite{PhysRevB.98.155438} for normal contacts 
($p_{2}^{(0)} = 0$). The new contributions in the NS case, second and third terms, are related to the AR and the coexistence 
of the two processes, respectively. The sign of $C_{\rm neq}^{\rm NS}$ depends on the precise energy dependence of 
$p^{(0)}_{1,2}$. One can gain some intuition by analyzing an NS contact with an energy-independent normal transmission coefficient.
In this case the $p^{(0)}_{1,2}$ coefficients can be computed e.g.\ with the well-known BTK model \cite{PhysRevB.25.4515}. 
As we show in Ref.~\cite{SM}, such an analysis suggests that two requirements must be simultaneously met to significantly 
break TUR: (i) coexistence of the two processes at a given bias and temperature and (ii) high transmissions. The coexistence
of the two processes is responsible for a large negative contribution of the cross term $-48p_1^{(0)}p_2^{(0)}$ in Eq.~(\ref{CneqNS}), 
while high transmissions are necessary to have high probabilities and make the terms $n^4p_n^{(0)}(1-6p_n^{(0)})$ negative. 
The dependence of the zero-bias probabilities on transmission and temperature, \textcolor{black}{as well as a discussion on the 
role of the different terms in Eq.~(\ref{CneqNS}), can be found in Ref.~\cite{SM}.} This intuition will be further confirmed 
below by our numerically exact results at arbitrary bias and we shall use those results to test the validity of Eq.~(\ref{CneqNS}). 

In the case of SS contacts, new transport processes in the form of multiple Andreev reflections (MAR) can take place. 
Their corresponding probabilities $p_{\pm n}$ for $n>2$ contribute to the current and noise (see 
Ref.~\cite{Cuevas2003,Cuevas2004}) and therefore appear in Eq.~(\ref{deviation}). The lowest order transport processes 
in an SS junction are depicted in Fig.~\ref{fig-1}(c-f) and correspond to QP ($p_{\pm 1}$), AR ($p_{\pm 2}$), MAR 
($p_{\pm 3}$) and 4th order MAR ($p_{\pm 4}$) transferring one, two, three and four electrons respectively.

\emph{QPC of constant transmission}.--- In the following we focus on single-channel NS and SS contacts characterized 
by an energy-independent normal state transmission coefficient $\tau$. These junctions have been realized with the help 
of superconducting atomic-size contacts in which both the current and the noise have been thoroughly studied 
\cite{Scheer1997,Scheer1998,Cuevas1998,Cron2001,Senkpiel2020,Senkpiel2020a}. We make use of the framework of full counting
 statistics to obtain current, noise and charge resolved probabilities using Ref.~\cite{Ohnmacht2023} to compute the full 
 result for the TUR-breaking coefficient in Eq.~(\ref{TURcoeff}). 

\begin{figure*}
    \centering
    \includegraphics[width=0.8\textwidth,clip]{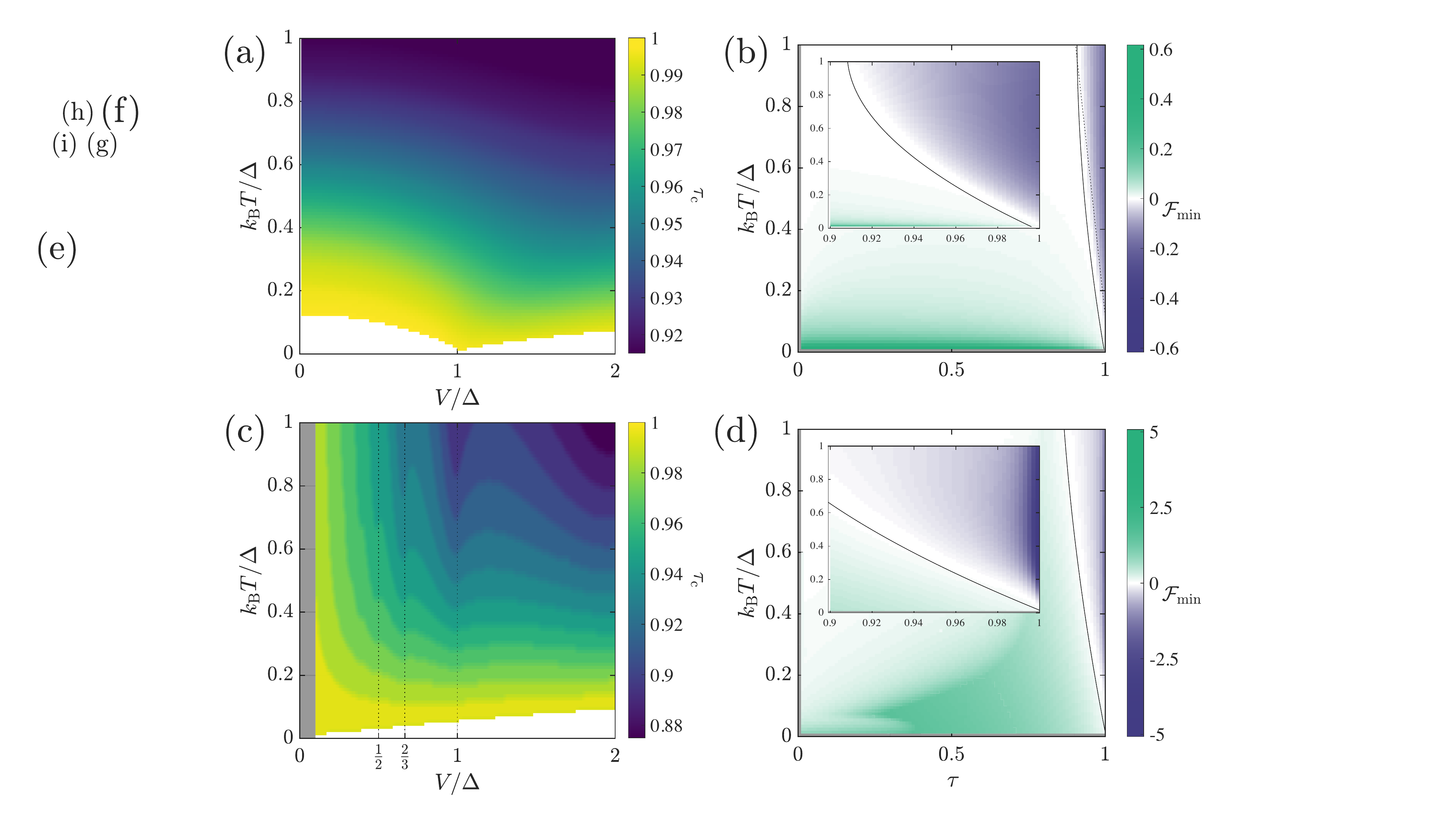}
    \caption{(a) Critical transmission $\tau_c$ for breaking TUR for a NS junction as a function of the bias voltage 
    $V/\Delta$ and the temperature $k_{\rm B}T/\Delta$. For areas with no color, TUR is not broken. (b) The minimal 
    TUR breaking coefficient $\mathcal{F}_{\rm min}$ as a function of the dimensionless temperature $k_{\rm B}T/\Delta$ 
    and the transmission $\tau$ for a NS junction. A negative coefficient indicates broken TUR. The black solid line 
    indicates the phase boundary. The dotted line indicates the approximate phase boundary from the coefficient 
    $C_{\rm neq}$ [see Eq.~(\ref{CneqNS})]. {\color{black} The inset shows a zoom-in into the region where TUR is broken.} (c) Same as in (a) for an SS junction. Dotted lines indicate special 
    voltages of onsets of different transport processes, namely $V/\Delta = 1, 2/3, 1/2$ for AR, MAR and 4th order MAR. 
    Grey area shows a parameter realm that was not analyzed. (d) Same as in (b) for an SS junction with the phase 
    boundary as a solid line. {\color{black} The inset shows a zoom-in into the region where TUR is broken.}}
    \label{fig-2}
\end{figure*}

Our purpose is to investigate whether TUR can be broken in a NS and SS system and portray the ``phase" diagram of 
the system. We assign the phase of unbroken TUR if $\mathcal{F}\geq  0$ and the second phase of broken TUR if 
$\mathcal{F}< 0$. To illustrate the results, we fix the voltage and temperature and vary the transmission $\tau$ of 
the QPC. We define $\tau_{\rm c}$ as the critical transmission for which the lower bound of TUR 
inequality is satisfied ($\mathcal{F}=0$). It holds that for any $\tau < \tau_{\rm c}$, TUR is fulfilled whereas it is 
violated for $\tau > \tau_{\rm c}$. In Fig.~\ref{fig-2}(a) we show $\tau_{\rm c}$ as a function of 
the dimensionless temperature $k_{\rm B}T/\Delta$ and voltage $V/\Delta$ for the NS case. Notice that the TUR only breaks
at high transmissions ($\tau_c \approx 0.91$), as anticipated in the discussion of Eq.~\ref{CneqNS}. Note also that
$\tau_{\rm c}$ decreases upon increasing the temperature, which is due to the larger overlap between the two 
process induced by the finite temperature. We want to 
stress that no energy dependence of the transmission is needed for breaking TUR. The source of nonlinearity of the
transport characteristics ($I$ and $S$) as a function of the bias that is required to violate TUR arises from the 
coexistence of QP tunneling and AR in addition to the non-constant density of states of the SC. This is in contrast 
to the results in Ref.~\cite{PhysRevB.98.155438}, where an energy dependence of the transmission is needed. For low
temperatures, TUR is only broken for $V \approx \Delta$ and only for $\tau \approx 1$. This indicates that the breaking 
of TUR is rooted in the coexistence of the QP and AR (for more details see Ref.~\cite{SM}).

To quantify the violation of TUR, we consider the voltage at which TUR is maximally broken and define the minimal 
TUR-breaking coefficient $\mathcal{F}_{\rm min} \equiv {\rm min}_{V} \mathcal{F}$, which is positive (or zero) for 
unbroken TUR and negative for broken TUR. In addition, the more negative $\mathcal{F}_{\rm min}$ the more TUR is 
broken. In Fig.~\ref{fig-2}(b), this coefficient is plotted as a function of the transmission and the temperature. 
The solid line indicates the phase boundary between the unbroken and broken TUR phase. Notice that the TUR is more 
broken the higher the transmission is. These findings can be understood with the help of the low-bias
analysis above, see Eq.~(\ref{CneqNS}). The probabilities $p_{1,2}$ are analytical in the NS (see 
Ref.~\cite{Ohnmacht2023,SM}) and $C_{\rm neq}^{\rm NS}$ can be easily calculated. The phase boundary obtained with 
this approximation is shown as a dotted line in panel Fig.~\ref{fig-2}(b). It qualitatively reproduces the exact results,
especially at high temperatures where the energy dependence of $p_{1,2}$ is mainly determined by the smearing of the
Fermi functions (for more details see Ref.~\cite{SM}). On the other hand, the fact that high transmissions are required
to have a stronger violation of TUR can be understood from the contribution $2^4p^{(0)}_2(1-6p^{(0)}_2)$ in Eq.~(\ref{CneqNS}). 
This term scales quartically with the charge, which renders a stronger violation for highly probable AR processes coexisting 
with low probability QP processes. Furthermore, an important contribution to $C_{\rm neq}^{\rm NS}$ 
comes also from the cross term $-48p_1^{(0)}p_2^{(0)}$.

For completeness, we also address in Ref.~\cite{SM} the case of an NS contact featuring an 
energy dependent transmission coefficient. In particular, we have studied the case in which the leads are coupled via 
a quantum dot. The transmission coefficient in this case corresponds to a Breit-Wigner resonance centered in the energy
of the dot level $\epsilon_0$ with a tunneling broadening given by $\Gamma$, i.e., $\tau(E)=\Gamma^2/[(E-\epsilon_0)^2+
\Gamma^2]$. However, we find that this added energy dependence does not further break TUR, see Ref.~\cite{SM}, which we 
attribute to the fact that the resonant level tends to reduce the high effective transmission at some energies that is 
needed to break TUR.

We turn now to the analysis of the SS case, where the presence of MAR processes, which are shown in Fig.~\ref{fig-1}(e-f), 
adds additional complexity. Again we refer to Ref.~\cite{Cuevas2003,Cuevas2004,Ohnmacht2023} for the current and noise 
calculation in the framework of the full counting statistics. The critical transmission $\tau_{\rm c}$ for the SS case 
is shown in Fig.~\ref{fig-2}(c). It is noticeable that the TUR is broken for generally smaller transmissions in comparison
to the NS case [see panel (a)]. For higher temperatures, the critical transmission decreases because the 
processes overlap more, which benefits the TUR breaking. As compared to the NS case, the critical transmission depends more 
drastically on the voltage even for higher temperatures. In particular, the dependence of the voltage is more pronounced 
in the vicinity of voltages corresponding to onsets of MAR processes, namely $V=2\Delta/n$, with $n$ being an integer. 
Also notice that for low temperatures, the TUR violation concentrates on only one point, namely $V = T = 0$. This can be 
explained by considering that in the limit $V,T\to 0$ for an SS junction with high transmission, all transmission 
probabilities up to infinite order contribute. In the case of perfect transmission, the current does not go to zero for 
$V,T\to 0$ and converges to a non-zero value \cite{Cuevas2003,Cuevas2004}, rendering the absolute limit of violation.

The minimal TUR breaking coefficient is displayed in Fig.~\ref{fig-2}(d), where it is seen that TUR is severely more 
broken than in the NS case, reaching absolute values of one order of magnitude higher than in the NS case. This enhanced 
breaking must originate from the MARs, which are exclusive to the SS. In particular, this 
large breaking results from highly probable, high-order MAR processes coexisting with lots of lower probability MARs of lower 
order, granting an immensely strong violation of the TUR. As the voltage range is restricted for $\mathcal{F}_{\rm min}$ in 
Fig.~\ref{fig-2}(d) due to computation time, $V \geq 0.1\Delta$, an even stronger violation is achievable for lower voltages.

\emph{Conclusions.--} In this work, we have studied the violation of TUR in the coherent charge transport in junctions
containing superconductors. We have shown that the presence of multiple tunneling processes help breaking TUR 
and that the degree of violation is clearly larger in the SS case. We have rationalized this by providing an analytical 
formula for the TUR-breaking coefficient for low voltages and show that it only depends on the zero-bias tunneling 
probabilities. We have also shown that for junctions containing superconductors, no energy dependence of the normal 
transmission coefficient is required to break TUR, \textcolor{black}{which is due to the intrinsic energy dependence of the 
probability of the different superconducting tunneling processes. This is a key advantage that makes it much more feasible to
observe the violation of TUR in superconducting junctions. So, in summary,} our work shows that superconducting junctions 
of simple quantum point contacts are promising candidates for the engineering of quantum thermal machines that are immune 
against fluctuations in a low dissipation scenario. Our contribution also paves the way for the experimental observation 
of the TUR breaking in the context of quantum electron transport, something that continues to be very elusive. 

D.C.O.\ and W.B.\ acknowledge support by the Deutsche Forschungsgemeinschaft (DFG; German Research Foundation) via SFB 1432
(Project No.\ 425217212). J.C.C.\ acknowledges support from the Spanish Ministry of Science and Innovation under Grant 
No.\ PID2020-114880GB-I00 and through the ``Mar\'{\i}a de Maeztu” Programme for Units of Excellence in R\&D (CEX2023-001316-M)
and thanks the DFG and SFB 1432 for sponsoring his stay at the University of Konstanz as a Mercator Fellow. R.L.\ acknowledges
support by the Spanish State Research Agency (MCIN/AEI/10.13039/501100011033) and FEDER (UE) under Grant No.\
PID2020-117347GB-I00 and María de Maeztu Project No.\ CEX2021-001164-M and Grant No.\ PDR2020/12 sponsored by the Comunitat 
Autonoma de les Illes Balears through the Direcció General de Política Universitaria i Recerca with funds from the Tourist 
Stay Tax Law ITS 2017-006. K.W.K.\ acknowledges the support by the National Research Foundation of Korea (NRF) grant 
funded by the Korea government (MSIT) (No.\ 2020R1A5A1016518).

\bibliography{MyLibrary_revised.bib}

@article{PhysRevB.108.115422,
  title = {Thermodynamic uncertainty relations for systems with broken time reversal symmetry: The case of superconducting hybrid systems},
  author = {Taddei, Fabio and Fazio, Rosario},
  journal = {Phys. Rev. B},
  volume = {108},
  issue = {11},
  pages = {115422},
  numpages = {7},
  year = {2023},
  month = {Sep},
  publisher = {American Physical Society},
  doi = {10.1103/PhysRevB.108.115422},
  url = {https://link.aps.org/doi/10.1103/PhysRevB.108.115422}
}

@article{nahum1994electronic,
  title = {Electronic Microrefrigerator Based on a Normal-insulator-superconductor Tunnel Junction},
  author = {Nahum, M. and Eiles, T. M. and Martinis, John M.},
  year = {1994},
  month = dec,
  journal = {, Appl. Phys. Lett.},
  volume = {65},
  number = {24},
  pages = {3123--3125},
  issn = {0003-6951},
  doi = {10.1063/1.112456},
  urldate = {2024-06-12}
}

@article{leivo1996efficient,
  title={Efficient Peltier refrigeration by a pair of normal metal/insulator/superconductor junctions},
  author={Leivo, M. M. and Pekola, J. P. and Averin, D. V.},
  journal={, Appl. Phys. Lett.},
  volume={68},
  number={14},
  pages={1996--1998},
  month={April},
  year={1996},
  doi={10.1063/1.115651},
  url={https://doi.org/10.1063/1.115651}
}

@article{PhysRevB.98.241414,
  title = {Cooling by Cooper pair splitting},
  author = {S\'anchez, Rafael and Burset, Pablo and Yeyati, Alfredo Levy},
  journal = {Phys. Rev. B},
  volume = {98},
  issue = {24},
  pages = {241414(R)},
  numpages = {6},
  year = {2018},
  month = {Dec},
  publisher = {American Physical Society},
  doi = {10.1103/PhysRevB.98.241414},
  url = {https://link.aps.org/doi/10.1103/PhysRevB.98.241414}
}

@article{PhysRevB.107.245412,
  title = {Superconductor--quantum dot hybrid coolers},
  author = {Hwang, Sun-Yong and Sothmann, Bj\"orn and S\'anchez, David},
  journal = {Phys. Rev. B},
  volume = {107},
  issue = {24},
  pages = {245412},
  numpages = {9},
  year = {2023},
  month = {Jun},
  publisher = {American Physical Society},
  doi = {10.1103/PhysRevB.107.245412},
  url = {https://link.aps.org/doi/10.1103/PhysRevB.107.245412}
}

@article{PhysRevResearch.5.043041,
  title = {Quantum-enhanced performance in superconducting Andreev reflection engines},
  author = {Manzano, Gonzalo and L\'opez, Rosa},
  journal = {Phys. Rev. Res.},
  volume = {5},
  issue = {4},
  pages = {043041},
  numpages = {10},
  year = {2023},
  month = {Oct},
  publisher = {American Physical Society},
  doi = {10.1103/PhysRevResearch.5.043041},
  url = {https://link.aps.org/doi/10.1103/PhysRevResearch.5.043041}
}

@article{PhysRevB.98.155438,
  title = {Assessing the validity of the thermodynamic uncertainty relation in quantum systems},
  author = {Agarwalla, Bijay Kumar and Segal, Dvira},
  journal = {Phys. Rev. B},
  volume = {98},
  issue = {15},
  pages = {155438},
  numpages = {9},
  year = {2018},
  month = {Oct},
  publisher = {American Physical Society},
  doi = {10.1103/PhysRevB.98.155438},
  url = {https://link.aps.org/doi/10.1103/PhysRevB.98.155438}
}

@article{PhysRevE.99.062141,
  title = {Thermodynamic uncertainty relation in quantum thermoelectric junctions},
  author = {Liu, Junjie and Segal, Dvira},
  journal = {Phys. Rev. E},
  volume = {99},
  issue = {6},
  pages = {062141},
  numpages = {12},
  year = {2019},
  month = {Jun},
  publisher = {American Physical Society},
  doi = {10.1103/PhysRevE.99.062141},
  url = {https://link.aps.org/doi/10.1103/PhysRevE.99.062141}
}

@article{PhysRevB.98.085425,
  title = {Coherence-enhanced constancy of a quantum thermoelectric generator},
  author = {Ptaszy\ifmmode \acute{n}\else \'{n}\fi{}ski, Krzysztof},
  journal = {Phys. Rev. B},
  volume = {98},
  issue = {8},
  pages = {085425},
  numpages = {11},
  year = {2018},
  month = {Aug},
  publisher = {American Physical Society},
  doi = {10.1103/PhysRevB.98.085425},
  url = {https://link.aps.org/doi/10.1103/PhysRevB.98.085425}
}

@article{PhysRevLett.120.090601,
  title = {Thermodynamic Bounds on Precision in Ballistic Multiterminal Transport},
  author = {Brandner, Kay and Hanazato, Taro and Saito, Keiji},
  journal = {Phys. Rev. Lett.},
  volume = {120},
  issue = {9},
  pages = {090601},
  numpages = {6},
  year = {2018},
  month = {Mar},
  publisher = {American Physical Society},
  doi = {10.1103/PhysRevLett.120.090601},
  url = {https://link.aps.org/doi/10.1103/PhysRevLett.120.090601}
}

@article{PhysRevX.11.021013,
  title = {Thermodynamic bounds on coherent transport in periodically driven conductors},
  author = {Potanina, Elina and Flindt, Christian and Moskalets, Michael and Brandner, Kay},
  journal = {Phys. Rev. X},
  volume = {11},
  issue = {2},
  pages = {021013},
  numpages = {26},
  year = {2021},
  month = {Apr},
  publisher = {American Physical Society},
  doi = {10.1103/PhysRevX.11.021013},
  url = {https://link.aps.org/doi/10.1103/PhysRevX.11.021013}
}

@article{PhysRevB.106.115419,
  title = {Nonlocal quantum heat engines made of hybrid superconducting devices},
  author = {Tabatabaei, S. Mojtaba and S\'anchez, David and Yeyati, Alfredo Levy and S\'anchez, Rafael},
  journal = {Phys. Rev. B},
  volume = {106},
  issue = {11},
  pages = {115419},
  numpages = {13},
  year = {2022},
  month = {Sep},
  publisher = {American Physical Society},
  doi = {10.1103/PhysRevB.106.115419},
  url = {https://link.aps.org/doi/10.1103/PhysRevB.106.115419}
}

@article{PhysRevB.103.235434,
  title = {Nonlocal thermoelectric engines in hybrid topological Josephson junctions},
  author = {Blasi, Gianmichele and Taddei, Fabio and Arrachea, Liliana and Carrega, Matteo and Braggio, Alessandro},
  journal = {Phys. Rev. B},
  volume = {103},
  issue = {23},
  pages = {235434},
  numpages = {16},
  year = {2021},
  month = {Jun},
  publisher = {American Physical Society},
  doi = {10.1103/PhysRevB.103.235434},
  url = {https://link.aps.org/doi/10.1103/PhysRevB.103.235434}
}

@article{PhysRevLett.124.106801,
  title = {Nonlinear Thermoelectricity with Electron-Hole Symmetric Systems},
  author = {Marchegiani, G. and Braggio, A. and Giazotto, F.},
  journal = {Phys. Rev. Lett.},
  volume = {124},
  issue = {10},
  pages = {106801},
  numpages = {6},
  year = {2020},
  month = {Mar},
  publisher = {American Physical Society},
  doi = {10.1103/PhysRevLett.124.106801},
  url = {https://link.aps.org/doi/10.1103/PhysRevLett.124.106801}
}

@article{germanese2022bipolar,
  title = {Bipolar Thermoelectric {{Josephson}} Engine},
  author = {Germanese, Gaia and Paolucci, Federico and Marchegiani, Giampiero and Braggio, Alessandro and Giazotto, Francesco},
  year = {2022},
  month = oct,
  journal = {Nat. Nanotechnol.},
  volume = {17},
  number = {10},
  pages = {1084--1090},
  publisher = {Nature Publishing Group},
  issn = {1748-3395},
  doi = {10.1038/s41565-022-01208-y},
  urldate = {2024-06-12},
  copyright = {2022 The Author(s), under exclusive licence to Springer Nature Limited},
  langid = {english}
}

@article{PhysRevResearch.6.L012022,
  title = {Bipolar thermoelectric superconducting single-electron transistor},
  author = {Battisti, Sebastiano and De Simoni, Giorgio and Chirolli, Luca and Braggio, Alessandro and Giazotto, Francesco},
  journal = {Phys. Rev. Res.},
  volume = {6},
  issue = {1},
  pages = {L012022},
  numpages = {6},
  year = {2024},
  month = {Jan},
  publisher = {American Physical Society},
  doi = {10.1103/PhysRevResearch.6.L012022},
  url = {https://link.aps.org/doi/10.1103/PhysRevResearch.6.L012022}
}

@article{PhysRevB.106.245413,
  title = {Voltage-amplified heat rectification in SIS junctions},
  author = {Khomchenko, Ilia and Ouerdane, Henni and Benenti, Giuliano},
  journal = {Phys. Rev. B},
  volume = {106},
  issue = {24},
  pages = {245413},
  numpages = {6},
  year = {2022},
  month = {Dec},
  publisher = {American Physical Society},
  doi = {10.1103/PhysRevB.106.245413},
  url = {https://link.aps.org/doi/10.1103/PhysRevB.106.245413}
}

@article{PhysRevB.104.045424,
  title = {Broadband frequency filters with quantum dot chains},
  author = {Ehrlich, Tilmann and Schaller, Gernot},
  journal = {Phys. Rev. B},
  volume = {104},
  issue = {4},
  pages = {045424},
  numpages = {12},
  year = {2021},
  month = {Jul},
  publisher = {American Physical Society},
  doi = {10.1103/PhysRevB.104.045424},
  url = {https://link.aps.org/doi/10.1103/PhysRevB.104.045424}
}

@article{Barato2015,
  title = {Thermodynamic Uncertainty Relation for Biomolecular Processes},
  author = {Barato, Andre C. and Seifert, Udo},
  journal = {Phys. Rev. Lett.},
  volume = {114},
  issue = {15},
  pages = {158101},
  numpages = {5},
  year = {2015},
  month = {Apr},
  publisher = {American Physical Society},
  doi = {10.1103/PhysRevLett.114.158101},
  url = {https://link.aps.org/doi/10.1103/PhysRevLett.114.158101}
}

@article{Todd2016,
  title = {Dissipation Bounds All Steady-State Fluctuations},
  author = {Gingrich, Todd R. and Horowitz, Jordan M. and Perunov, Nikolay and England, Jeremy L.},
  journal = {Phys. Rev. Lett.},
  volume = {116},
  issue = {12},
  pages = {120601},
  numpages = {5},
  year = {2016},
  month = {Mar},
  publisher = {American Physical Society},
  doi = {10.1103/PhysRevLett.116.120601},
  url = {https://link.aps.org/doi/10.1103/PhysRevLett.116.120601}
}

@article{Horowitz2020,
  title = {Thermodynamic uncertainty relations constrain non-equilibrium fluctuations},
  author = {Horowitz, Jordan M. and Gingrich, Todd R.},
  journal = {Nat. Phys.},
  volume = {16},
  pages = {15-20},
  numpages = {6},
  year = {2020},
  doi = {10.1038/s41567-019-0702-6},
  url = {https://doi.org/10.1038/s41567-019-0702-6}
}

@article{Pietz2016,
  title = {Universal bounds on current fluctuations},
  author = {Pietzonka, Patrick and Barato, Andre C. and Seifert, Udo},
  journal = {Phys. Rev. E},
  volume = {93},
  issue = {5},
  pages = {052145},
  numpages = {16},
  year = {2016},
  month = {May},
  publisher = {American Physical Society},
  doi = {10.1103/PhysRevE.93.052145},
  url = {https://link.aps.org/doi/10.1103/PhysRevE.93.052145}
}

@article{Pietzonka16b,
doi = {10.1088/1742-5468/2016/12/124004},
url = {https://dx.doi.org/10.1088/1742-5468/2016/12/124004},
year = {2016},
month = {dec},
publisher = {IOP Publishing and SISSA},
volume = {2016},
number = {12},
pages = {124004},
author = {Patrick Pietzonka and Andre C Barato and Udo Seifert},
title = {Universal bound on the efficiency of molecular motors},
journal = {Journal of Statistical Mechanics: Theory and Experiment},
}

@article{Pietz2018,
  title = {Universal Trade-Off between Power, Efficiency, and Constancy in Steady-State Heat Engines},
  author = {Pietzonka, Patrick and Seifert, Udo},
  journal = {Phys. Rev. Lett.},
  volume = {120},
  issue = {19},
  pages = {190602},
  numpages = {6},
  year = {2018},
  month = {May},
  publisher = {American Physical Society},
  doi = {10.1103/PhysRevLett.120.190602},
  url = {https://link.aps.org/doi/10.1103/PhysRevLett.120.190602}
}

@article{Timp2019,
  title = {Thermodynamic Uncertainty Relations from Exchange Fluctuation Theorems},
  author = {Timpanaro, Andr\'e M. and Guarnieri, Giacomo and Goold, John and Landi, Gabriel T.},
  journal = {Phys. Rev. Lett.},
  volume = {123},
  issue = {9},
  pages = {090604},
  numpages = {6},
  year = {2019},
  month = {Aug},
  publisher = {American Physical Society},
}

@article{PhysRevLett.101.136805,
  title = {Fluctuation Relations Without Microreversibility in Nonlinear Transport},
  author = {F\"orster, H. and B\"uttiker, M.},
  journal = {Phys. Rev. Lett.},
  volume = {101},
  issue = {13},
  pages = {136805},
  numpages = {4},
  year = {2008},
  month = {Sep},
  publisher = {American Physical Society}
}

@article{PhysRevB.72.235328,
  title = {Inelastic interaction corrections and universal relations for full counting statistics in a quantum contact},
  author = {Tobiska, J. and Nazarov, Yu. V.},
  journal = {Phys. Rev. B},
  volume = {72},
  issue = {23},
  pages = {235328},
  numpages = {10},
  year = {2005},
  month = {Dec},
  publisher = {American Physical Society},

}

@article{Andrieux_2007,
doi = {10.1088/1742-5468/2007/02/P02006},
url = {https://dx.doi.org/10.1088/1742-5468/2007/02/P02006},
year = {2007},
month = {feb},
publisher = {},
volume = {2007},
number = {02},
pages = {P02006},
author = {David Andrieux and Pierre Gaspard},
title = {A fluctuation theorem for currents and non-linear response coefficients},
journal = {Journal of Statistical Mechanics: Theory and Experiment},
}

@article{PhysRevLett.108.246603,
  title = {Fluctuation Relations for Spintronics},
  author = {L\'opez, Rosa and Lim, Jong Soo and S\'anchez, David},
  journal = {Phys. Rev. Lett.},
  volume = {108},
  issue = {24},
  pages = {246603},
  numpages = {5},
  year = {2012},
  month = {Jun},
  publisher = {American Physical Society},
  doi = {10.1103/PhysRevLett.108.246603},
  url = {https://link.aps.org/doi/10.1103/PhysRevLett.108.246603}
}

@article{Cuevas2003,
  title = {Full {{Counting Statistics}} of {{Multiple Andreev Reflections}}},
  author = {Cuevas, J. C. and Belzig, W.},
  year = {2003},
  month = oct,
  journal = {Phys. Rev. Lett.},
  volume = {91},
  number = {18},
  pages = {187001},
  publisher = {American Physical Society},
  doi = {10.1103/PhysRevLett.91.187001},
  urldate = {2024-05-08}
}

@article{Cuevas2004,
  title = {Dc Transport in Superconducting Point Contacts:  {{A}} Full-Counting-Statistics View},
  shorttitle = {Dc Transport in Superconducting Point Contacts},
  author = {Cuevas, J. C. and Belzig, W.},
  year = {2004},
  month = dec,
  journal = {Phys. Rev. B},
  volume = {70},
  number = {21},
  pages = {214512},
  publisher = {American Physical Society},
  doi = {10.1103/PhysRevB.70.214512},
  urldate = {2024-05-08}
}

@article{Levitov1996,
  title = {Electron Counting Statistics and Coherent States of Electric Current},
  author = {Levitov, Leonid S. and Lee, Hyunwoo and Lesovik, Gordey B.},
  year = {1996},
  month = oct,
  journal = {Journal of Mathematical Physics},
  volume = {37},
  number = {10},
  pages = {4845--4866},
  issn = {0022-2488, 1089-7658},
  doi = {10.1063/1.531672},
  urldate = {2024-05-08},
  langid = {english}
}

@article{PhysRevResearch.5.023155,
  title = {Entanglement and thermokinetic uncertainty relations in coherent mesoscopic transport},
  author = {Prech, Kacper and Johansson, Philip and Nyholm, Elias and Landi, Gabriel T. and Verdozzi, Claudio and Samuelsson, Peter and Potts, Patrick P.},
  journal = {Phys. Rev. Res.},
  volume = {5},
  issue = {2},
  pages = {023155},
  numpages = {22},
  year = {2023},
  month = {Jun},
  publisher = {American Physical Society},
  doi = {10.1103/PhysRevResearch.5.023155},
  url = {https://link.aps.org/doi/10.1103/PhysRevResearch.5.023155}
}

@article{PhysRevB.28.1655,
  title = {Phase transitions in stationary nonequilibrium states of model lattice systems},
  author = {Katz, Sheldon and Lebowitz, Joel L. and Spohn, H.},
  journal = {Phys. Rev. B},
  volume = {28},
  issue = {3},
  pages = {1655--1658},
  numpages = {0},
  year = {1983},
  month = {Aug},
  publisher = {American Physical Society},
  doi = {10.1103/PhysRevB.28.1655},
  url = {https://link.aps.org/doi/10.1103/PhysRevB.28.1655}
}

@article{PhysRevResearch.5.013038,
  title = {Optimal superconducting hybrid machine},
  author = {L\'opez, Rosa and Lim, Jong Soo and Kim, Kun Woo},
  journal = {Phys. Rev. Res.},
  volume = {5},
  issue = {1},
  pages = {013038},
  numpages = {10},
  year = {2023},
  month = {Jan},
  publisher = {American Physical Society},
  doi = {10.1103/PhysRevResearch.5.013038},
  url = {https://link.aps.org/doi/10.1103/PhysRevResearch.5.013038}
}

@article{Ohnmacht2023,
  title = {Full Counting Statistics of {{Yu-Shiba-Rusinov}} Bound States},
  author = {Ohnmacht, David Christian and Belzig, Wolfgang and Cuevas, Juan Carlos},
  year = {2023},
  month = sep,
  journal = {Phys. Rev. Res.},
  volume = {5},
  number = {3},
  pages = {033176},
  publisher = {American Physical Society},
  doi = {10.1103/PhysRevResearch.5.033176},
  urldate = {2024-05-08}
}

@article{Scheer1997,
  title = {Conduction {{Channel Transmissions}} of {{Atomic-Size Aluminum Contacts}}},
  author = {Scheer, E. and Joyez, P. and Esteve, D. and Urbina, C. and Devoret, M. H.},
  year = {1997},
  month = may,
  journal = {Phys. Rev. Lett.},
  volume = {78},
  number = {18},
  pages = {3535--3538},
  publisher = {American Physical Society},
  doi = {10.1103/PhysRevLett.78.3535},
  urldate = {2024-06-03}
}

@article{Senkpiel2020,
  title = {Single Channel {{Josephson}} Effect in a High Transmission Atomic Contact},
  author = {Senkpiel, Jacob and Dambach, Simon and Etzkorn, Markus and Drost, Robert and Padurariu, Ciprian and Kubala, Bj{\"o}rn and Belzig, Wolfgang and Yeyati, Alfredo Levy and Cuevas, Juan Carlos and Ankerhold, Joachim and Ast, Christian R. and Kern, Klaus},
  year = {2020},
  month = jul,
  journal = {Commun. Phys.},
  volume = {3},
  pages = {131},
  publisher = {Nature Publishing Group},
  issn = {2399-3650},
  doi = {10.1038/s42005-020-00397-z},
  urldate = {2024-06-03},
  copyright = {2020 The Author(s)},
  langid = {english}
}

@article{Cuevas1998,
  title = {Evolution of {{Conducting Channels}} in {{Metallic Atomic Contacts}} under {{Elastic Deformation}}},
  author = {Cuevas, J. C. and Levy Yeyati, A. and {Mart{\'i}n-Rodero}, A. and Bollinger, G. R. and Untiedt, C. and Agra{\"i}t, N.},
  year = {1998},
  month = oct,
  journal = {Phys. Rev. Lett.},
  volume = {81},
  number = {14},
  pages = {2990--2993},
  publisher = {American Physical Society},
  doi = {10.1103/PhysRevLett.81.2990},
  urldate = {2024-06-03}
}

@article{Scheer1998,
  title = {The Signature of Chemical Valence in the Electrical Conduction through a Single-Atom Contact},
  author = {Scheer, Elke and Agra{\"i}t, Nicol{\'a}s and Cuevas, Juan Carlos and Yeyati, Alfredo Levy and Ludoph, Bas and {Mart{\'i}n-Rodero}, Alvaro and Bollinger, Gabino Rubio and {van Ruitenbeek}, Jan M. and Urbina, Cristi{\'a}n},
  year = {1998},
  month = jul,
  journal = {Nature},
  volume = {394},
  number = {6689},
  pages = {154--157},
  publisher = {Nature Publishing Group},
  issn = {1476-4687},
  doi = {10.1038/28112},
  urldate = {2024-06-03},
  copyright = {1998 Macmillan Magazines Ltd.},
  langid = {english}
}

@article{Senkpiel2020a,
  title = {Dynamical {{Coulomb Blockade}} as a {{Local Probe}} for {{Quantum Transport}}},
  author = {Senkpiel, Jacob and Kl{\"o}ckner, Jan C. and Etzkorn, Markus and Dambach, Simon and Kubala, Bj{\"o}rn and Belzig, Wolfgang and Yeyati, Alfredo Levy and Cuevas, Juan Carlos and Pauly, Fabian and Ankerhold, Joachim and Ast, Christian R. and Kern, Klaus},
  year = {2020},
  month = apr,
  journal = {Phys. Rev. Lett.},
  volume = {124},
  number = {15},
  pages = {156803},
  publisher = {American Physical Society},
  doi = {10.1103/PhysRevLett.124.156803},
  urldate = {2024-06-03}
}

@article{Cron2001,
  title = {Multiple-{{Charge-Quanta Shot Noise}} in {{Superconducting Atomic Contacts}}},
  author = {Cron, R. and Goffman, M. F. and Esteve, D. and Urbina, C.},
  year = {2001},
  month = apr,
  journal = {Phys. Rev. Lett.},
  volume = {86},
  number = {18},
  pages = {4104--4107},
  publisher = {American Physical Society},
  doi = {10.1103/PhysRevLett.86.4104},
  urldate = {2024-06-03}
}

@incollection{Belzig2003,
  title = {Full {{Counting Statistics}} of {{Superconductor-Normal-Metal Heterostructures}}},
  booktitle = {Quantum {{Noise}} in {{Mesoscopic Physics}}},
  author = {Belzig, W.},
  editor = {Nazarov, Yuli V.},
  year = {2003},
  pages = {463--496},
  publisher = {Springer Netherlands},
  address = {Dordrecht},
  doi = {10.1007/978-94-010-0089-5_22},
  urldate = {2024-06-03},
  isbn = {978-94-010-0089-5},
  langid = {english}
}

@note{SM,
  note = {See Supplemental Material for more detailed information on the charge-resolved probabilities.}
}

@incollection{Klich2003,
  title = {An {{Elementary Derivation}} of {{Levitov}}'s {{Formula}}},
  booktitle = {Quantum {{Noise}} in {{Mesoscopic Physics}}},
  author = {Klich, I.},
  editor = {Nazarov, Yuli V.},
  year = {2003},
  pages = {397--402},
  publisher = {Springer Netherlands},
  address = {Dordrecht},
  doi = {10.1007/978-94-010-0089-5_19},
  urldate = {2024-06-12},
  isbn = {978-94-010-0089-5},
  langid = {english}
}

@article{PhysRevB.101.195423,
  title = {Thermodynamic uncertainty relation in atomic-scale quantum conductors},
  author = {Friedman, Hava Meira and Agarwalla, Bijay K. and Shein-Lumbroso, Ofir and Tal, Oren and Segal, Dvira},
  journal = {Phys. Rev. B},
  volume = {101},
  issue = {19},
  pages = {195423},
  numpages = {9},
  year = {2020},
  month = {May},
  publisher = {American Physical Society},
  doi = {10.1103/PhysRevB.101.195423},
  url = {https://link.aps.org/doi/10.1103/PhysRevB.101.195423}
}

@article{PhysRevB.25.4515,
  title = {Transition from metallic to tunneling regimes in superconducting microconstrictions: Excess current, charge imbalance, and supercurrent conversion},
  author = {Blonder, G. E. and Tinkham, M. and Klapwijk, T. M.},
  journal = {Phys. Rev. B},
  volume = {25},
  issue = {7},
  pages = {4515--4532},
  numpages = {0},
  year = {1982},
  month = {Apr},
  publisher = {American Physical Society},
  doi = {10.1103/PhysRevB.25.4515},
  url = {https://link.aps.org/doi/10.1103/PhysRevB.25.4515}
}

\end{document}